\documentstyle[aas2pp4]{article}

\lefthead{Wiklind \& Combes}
\righthead{Complex molecular absorption at z$=$0.886}

\begin{document}

\title{The complex molecular absorption line system at z$=$0.886 towards
PKS1830--211}

\author{Tommy Wiklind}
\affil{Onsala Space Observatory, S--43992 Onsala, Sweden; 
(tommy@oso.chalmers.se)}

\and

\author{Fran\c{c}oise Combes}
\affil{DEMIRM, Observatoire de Paris, 61 Av. de l'Observatoire,
F--75014 Paris, France; (bottaro@obspm.fr)}

\begin{abstract}

New single dish and millimeter wave interferometer observations of the
molecular absorption line system in the gravitational lens to PKS1830--211
at z$=$0.88582 is presented.
Self--calibrated interferometer data shows unequivocally that the previously
detected absorption component is associated with the gravitationally lensed
south--west image of the background source. A second absorption line of
HCO$^+$(2$\leftarrow$1) at z$=$0.88582 is detected. This component is
shifted in velocity by --147\,km\,s$^{-1}$ relative to the main absorption
line, and is shown to be associated with the north--east image.
A few additional absorption lines are presented.
Upper limits to absorption and emission lines from the possible absorption 
system at z$=$0.1927, seen in 21\,cm HI by Lovell et al, are reported.
Alternatively, there could be only one absorbing system at z$=$0.88582,
with the HI line attributed instead to the PS molecule.

\end{abstract}


\section{Introduction}

The molecular interstellar medium (ISM) in galaxies at redshifts
up to z$\approx$0.9 has been studied through the use of absorption
of molecular rotational transitions (cf. Wiklind \& Combes 1996a,b;
Combes \& Wiklind 1995). Detection of several different
lines of molecules such as CO, HCO$^+$, HCN, HNC, CS,
N$_2$H$^+$, H$_2$CO, etc. allows a derivation of accurate
excitation temperatures, column densities and abundance ratios
(Wiklind \& Combes 1994, 1995, 1996a,b).
Four molecular absorption line systems are known to date;
in two of them the absorption occurs in the host galaxy to
the AGN and two represent truly intervening galaxies.
Since the distribution of molecular gas in galaxies is
usually centrally concentrated, the likelihood to encounter
a molecular cloud is higher if the line of sight passes close
to the center of an intervening galaxy. This means that
also the likelihood for gravitational lensing is high and,
indeed, in each of the two systems where absorption occurs
in an intervening galaxy, that galaxy acts as a gravitational
lens to the background source (Wiklind \& Combes 1995, 1996a)

The use of molecular absorption lines is manifold. Apart
from studying the physical and chemical properties of
molecular gas in distant galaxies, they can be used as
probes of very small scale structures in the molecular gas.
They can also be used as cosmographic probes, to set
upper limits to the temperature of the cosmic microwave
background radiation and to constrain the geometry in
gravitationally lensed systems.
The molecular ISM is probed on a scale of a few tenths of a
parsec, limited only by the angular extent of the background
source. The main uncertainty in the molecular gas properties
derived from the absorption lines is due to our ignorance of
possible small scale structures in the ISM on scales smaller
than the angular extent of the background source (e.g. Wiklind
\& Combes 1997).

In this article, we concentrate on the absorption line system
originating in a galaxy
which plays the role of a gravitational lens at z$=$0.886
(Wiklind \& Combes 1996a). The galaxy absorbs the continuum of
the strong flat--spectrum radio source PKS1830--211.
This radio source has long been considered a
gravitational lens candidate on basis of the
radio structure (Rao \& Subrahmanyan 1988;
Jauncey et al. 1991).
Since optical and near--infrared searches for
an optical counterpart to the radio source as well
to the lensing galaxy have remained unsuccessful
(cf. Djorgovski et al. 1992), the modeling of
the lensing system was lacking redshift information
(Kochanek \& Narayan 1992; Nair, Narasimha \& Rao 1993).
The galaxy at z$=$0.886 was discovered by
searching for molecular absorption lines in the
3-- and 2--mm band (Wiklind \& Combes 1996a).
A second intervening system has been reported at z$=$0.193,
seen through 21\,cm HI absorption by Lovell et al. (1996). 

We present new molecular absorption
data on the galaxy at z$=$0.886 as well as
upper limits to molecular absorption and emission
on the z$=$0.193 system. We also discuss what can be 
inferred regarding the mass of the lensing galaxy
and the redshift of the background source.

\section {Observations and data reduction}

Our data were obtained using the IRAM 
millimeter wave interferometer situated on Plateau de
Bure in France, the single dish IRAM 30--m telescope on Pico Veleta
in Spain and the 15--m SEST (Swedish ESO Submillimeter Telescope) on
La Silla in Chile.
The interferometer observations were done on several occasions
during 1995 and 1996. The single dish observations were likewise
obtained during several observing runs in 1995 and 1996.

At the IRAM 30--m telescope we used 3--, 2--, and 1.3--mm SIS
receivers tuned to the redshifted frequency of the observed
transition. The observations were made using a nutating subreflector
with a frequency of 0.5\,Hz and with a throw of $\pm$90'' in azimuth.
Typical system temperatures in the $T_{\rm A}^{*}$ scale were
400--600\,K. The weather conditions were good on most observing sessions.
Pointing, focusing and calibration were done before each observation.
As backends we used narrow band autocorrelators, giving a velocity
resolution of approximately 0.7\,km\,s$^{-1}$ and a bandwidth of
400\,km\,s$^{-1}$.
At the SEST we used 3--, and 2--mm SIS receivers with typical
system temperatures 200--300\,K. Here we used an AOS with a bandwidth
of 1200\,km\,s$^{-1}$ and a velocity resolution of 0.7\,km\,s$^{-1}$.
These observations were done in a dual beamswitch mode, where the
beam is displaced 12' in azimuth at a frequency of 6\,Hz.
The SEST data was obtained under good weather conditions. Pointing
and calibration where done as at IRAM.

The interferometer data were made with the standard BC configuration
(see Guilloteau et al. 1992). The array comprised 4 15--m telescopes.
The receivers were 3--mm SIS, giving a typical system temperature
of 400--500\,K. PKS1830--211 itself was used as a phase reference, 
while bandpass and amplitude calibrations were done using also
NRAO 530, 3C454.3 and W3OH.
The data reduction was made with the standard CLIC software.
Due to the low elevation of the PKS1830--211, we 
obtained a beam of 6 x 2.4 arcsec. This appears insufficient to resolve
two images separated by only one arcsec; but due to the high
signal--to--noise ratio, we were able to fit the integrated line
profile and the phase center, as a function of frequency, directly
from the visibilities.
We also made a two point sources fit, which gave the relative
positions of the two images and their relative fluxes.

The accuracy of the measurement depends on the inherent uncertainty in
the chopper--wheel calibration method as well as the pointing of the
telescope. The overall accuracy for a single observation on the
intensities are around 5--10\% for the interferometer data and
10--20\% for the single dish data.

\section{Results}

\subsection{The location of the z$=$0.886 absorption}

In our first study of the molecular absorption line
system towards PKS1830--211 we found that the
HCO$^+$(2$\leftarrow$1), HCN(2$\leftarrow$1) and,
possibly, the HNC(2$\leftarrow$1) lines were heavily
saturated (Wiklind \& Combes 1996a). Yet, the
depths of these absorption lines were only 36\%
of the total continuum. Since the magnification
ratio derived from long wavelength interferometer
data is 1.8 (Nair et al. 1993), corresponding
to a flux fraction of $\sim$36\% for the SW and
64\% for the NE core\footnote{Henceforth we refer
the the north--east component as NE and the
south--west component as SW.}, respectively,
we concluded that the molecular gas only covers the
SW core, with a high covering factor (close to 100\%). 
The present interferometric results confirm that 
only the SW component is absorbed around $V=0$.

In Fig.\,1 we show the HCO$^+$(2$\leftarrow$1)
spectrum obtained with the Plateau de Bure
interferometer.
The low declination of PKS1830--211 makes imaging
difficult, partly due to a north-south
elongated beam, and to the large air--mass.
The continuum is, however, strong
enough to allow a self--calibration of the data.
We made a gaussian fit of the 
source directly from the visibility data, and 
traced the location of the phase--center as a
function of frequency. The accuracy is very good
in RA ($\sim$0.05'') and slightly worse in DEC
($\sim$0.1'').
The two top panels in Fig.\,1 show the relative
phase in RA and DEC as a function of velocity across
the absorption profile. From the data obtained in 
September and October 1995, the maximum shifts are
$\Delta \alpha = 0.22''$ and $\Delta \delta = 0.28''$.
From the April 1996 data we get $\Delta \alpha = 0.24''$
and $\Delta \delta = 0.26''$ (see also Fig.\,3).

If the flux fraction from the NE and SW cores are
$f_1$ and $f_2$, respectively ($f_1 + f_2 = 1$),
the shift of the phase--center along the line
joining the cores is
\begin{eqnarray}
\Delta \alpha = \beta(v) f_2 r \sin{\psi} \ \ \ ,
\ \Delta \delta = \beta(v) f_2 r \cos{\psi} \ , \nonumber
\end{eqnarray}
where $\beta(v)$ is the fraction of continuum flux
from the SW core that is covered by optically thick
molecular gas of velocity $v$,
$r$ is the separation of the two components and
$\psi$ is the position angle of the line joining
the two cores, measured from north to east.
From the 15\,GHz VLA continuum map presented by
Subrahmanyan et al. (1990), we derive $r = 0.98''$
and $\psi = 44^{\circ}$.
Expressing the flux fractions in terms of the
magnification ratio $R = f_1/f_2$, we get
$\beta = (1+R) \Delta \alpha /(r \sin{\psi})$.
Using the magnification ratio $R$ derived from
long wavelength interferometry of 1.8 (Nair et al.
1993) and assuming that the covering factor $\beta$
is 100\% for the SW component, the phase--center
should shift $\Delta \alpha \approx 0.24''$ and
$\Delta \delta \approx 0.25''$, which is
consistent with the shift that we see in Fig. 1.

Frye, Welch \& Broadhurst (1997) observed PKS1830--211
with the BIMA interferometer and derived a
magnification ratio of $1.14 \pm 0.05$
by resolving the two continuum components.
Since the HCO$^+$(2$\leftarrow$1) line is manifestly
optically thick, as seen through the detection of the
H$^{13}$CO$^+$(2$\leftarrow$1) line (Wiklind \& Combes
1996a), such a low magnification ratio indicates that
the covering factor $\beta$ of the SW core is only 80--84\%.
However, they measure a total flux of 2.25 Jy in March 96
(1 March and 20 March),
much larger than ours of $1.55 \pm 0.06$ Jy on 28 February
and $1.45 \pm 0.08$ Jy on 14 April 96). Measurements
made with the Owens Valley interferometer on 1 March
gave $1.6 \pm 0.1$ Jy (S. Aalto priv. comm.). It thus
appears that the continuum is overestimated by a factor
1.4--1.5 in Frye et al. (1997).
We stress that calibration is quite delicate since there
exist significant rapid variations in the continuum levels
of the two images. 
Given the high signal--to--noise of the Plateau de Bure
observations, we have tried to retrieve the magnification
ratio from  a two point sources fit of the visibility data. 
The fits were very robust, completely independent of the initial
parameters, and revealed an RA and DEC separation between the two
images exactly consistent with $r=0.98"$ and $\psi=44^{\circ}$
in all the 136 channels without absorption.
In Table 1 we show the magnification ratios obtained at the
various epochs. Although the absolute calibration is not
so precise, we believe that the relative ratios between the two 
components is accurate.
Although the magnification ratio has changed from
$1.8 \pm 0.1$ to $1.2 \pm 0.1$ between October 1995
and April 1996, in concordance with the value derived
by Frye et al, the lower continuum level measured by us
is in agreement with a covering factor $\sim$1 for the
SW component (see Sect.\,4.2).

\subsection{A second absorbing cloud at z$=$0.88489}

We have detected a second absorption component situated
at a velocity $-$147\,km\,s$^{-1}$ relative to the main
HCO$^+$(2$\leftarrow$1) absorption at z$=$0.88582.
The second component is  hereafter refered to as the
`satellite'. The HCO$^+$(2$\leftarrow$1) line has
been observed extensively during a monitoring campaign
at both the IRAM 30--m and SEST 15--m telescope.
Averages of data obtained during 1996 are shown in
Fig.\,2, where the two dashed vertical lines mark zero
velocity and $-147$\,km\,s$^{-1}$ in a heliocentric
velocity scale in the restframe of the z$=$0.88582
absorber. The satellite line is not seen in the
HCO$^+$(3$\leftarrow$2) spectrum.
Frye et al (1997) could have detected this component
in March 96 (since the line was present at this epoch,
see Fig. 3), but their lower signal-to noise ratio 
prevented the detection.

Could the second absorption line be associated with
the z$=$0.193 HI absorber rather than the one at
z$=$0.886? The only line at z$=$0.193 which could be
in the `vicinity' of the observed frequency of the
satellite line at 94.635\,GHz is the redshifted
CO(1--0) line at 96.650\,GHz. The difference
is in excess of 2\,GHz, corresponding to a
velocity difference of $\sim$6200\,km\,s$^{-1}$
in the restframe of the z$=$0.193 absorber.
It is therefore highly unlikely that the satellite
absorption is associated with this system.
We conclude that the satellite line
is a second HCO$^+$(2$\leftarrow$1) line at a
blueshifted velocity of $-147$\,km\,s$^{-1}$
relative to the redshift z$=$0.88582 (corresponding
to a redshift of z$=0.884896 \pm 0.00001$). 

Whereas the main absorption line at $V=0$ has a depth of
38--40\% of the total continuum, the satellite
line has a depth of only 7\%. Measured in
antenna temperature, this corresponds to 1.5\,mK
and 6\,mK in the SEST and IRAM data, as of
April 96. The actual opacity, however,
depends on the surface covering factor of the
absorption.
The separation of the NE and SW continuum components
is 0.98'', corresponding to a distance of 5.6\,kpc,
measured in the plane of the sky
(H$_0 = 75$\,km\,s$^{-1}$\,Mpc$^{-1}$, q$_0$=0.5).
The satellite absorption can therefore be
associated with only one of the continuum sources.
Since the flux contribution from the NE and SW
components are $\sim$60\% and $\sim$40\%,
respectively, the satellite absorption line is
either optically thin or has a very small
covering factor.

The satellite absorption line is clearly seen
in the interferometer data (Fig.\,3).
The relative location of the phase--center
for $\Delta \alpha$ is shown in the top panel.
In addition to the positive shift of the phase--center
already established for the main absorption around
$V=0$\,km\,s$^{-1}$, a shift is also seen at the
velocity of the satellite line, but this time towards
negative values.
This shows that the phase--center moves partly
towards the SW continuum component at the
velocity of the satellite absorption.
The second absorption component must therefore
originate in front of the NE continuum core.
We are thus seeing molecular gas from two
different parts of the intervening galaxy.

Two absorption components, separated by
330\,km\,s$^{-1}$, are also seen
at z$=$0.673 towards the radio source B3\,1504+377
(Wiklind \& Combes 1996b). In this case the
absorption occurs in the galaxy hosting the
radio source. The galaxy is close to
edge--on and the separation of
the two absorption lines are caused by
non--circular velocities along a single
line of sight within the disk
of the host galaxy, most likely caused by
the presence of a barred potential.
The line of sight to the two images of
the continuum core of PKS1830-211
passes through the intervening galaxy at
$\sim$1.8 and $\sim$3.8\,kpc (SW and
NE components) from the lens center
(Nair et al. 1993). A normal rotation curve
(and circular orbits) are sufficient here to 
account for the observations.

We derived column densities using the formalism
presented in Wiklind \& Combes (1995, 1997).
The HCO$^+$ column density of the satellite
absorber, derived using 60\% of the total flux,
is only $7 \times 10^{12}$\,cm$^{-2}$ 
(however we can not exclude the case of a highly
saturated line with a small covering factor).
Using 40\% of the total flux for the main
line, we find an HCO$^+$ column density
$> 5 \times 10^{14}$\,cm$^{-2}$.
This value is derived from the averaged
monitoring data for 1996 and is slightly
higher than the column density reported
in Wiklind \& Combes (1996a).
Since the main line is heavily saturated,
the difference is due to the much better
signal--to--noise ratio of the new data.
The column densities have been derived
using an excitation temperature $T_{\rm x} = 6$\,K.
Integrating the J$=$3$\leftarrow$2 spectrum over
the extent of the J$=$2$\leftarrow$1 line sets a
3$\sigma$ upper limit to the excitation
temperature of 8\,K (a 2$\sigma$ limit implies
$T_{\rm x} < 6$\,K).

\subsection{The z$=$0.193 21\,cm HI absorber}

We searched for redshifted CO(1$\leftarrow$0) absorption
and/or emission as well as HCO$^+$(2$\leftarrow$1)
absorption at z$=$0.19267. Neither emission nor
absorption was found. The CO and HCO$^+$ spectra
are shown in Fig.\,4.

\subsubsection{Limits to the CO and HCO$^+$ column
densities}

The continuum level at the redshifted frequency of
the CO(1$\leftarrow$0) and HCO$^+$(2$\leftarrow$1)
lines are 50\,mK and 26\,mK, at the SEST telescope. 
However, when deriving
upper limits to the column density for molecular
absorption we have to take the separate flux contributions
into account. While the 21\,cm HI absorption only
occurs over the NE component (Lovell et al. 1996),
putative molecular absorption can occur in front of
both components. In Table\,2 we give upper limits
for three cases; (1) the total continuum flux, (2)
60\% of the flux, corresponding to the NE component,
and (3) 40\%, corresponding to the SW component.
In all cases we assume an excitation temperature
of 6\,K. The ratio of the upper limits of the
molecular column density to that of the HI is
done for the NE component, since this line of
sight causes the atomic absorption at z$=$0.193.

We derived the HI column density from the spectra
presented in Lovell et al. (1996), by fitting two
gaussian profiles. The resulting column density is
$N_{\rm HI} = 1.5 \times 10^{18} T_{\rm sp}/f$\,cm$^{-2}$,
where $T_{\rm sp}$ is the spin temperature and $f$ is
the area filling factor. Since the HI absorption
is only seen towards the NE component
$f = 0.6 f^{\prime}$, where the factor 0.6
corresponds to the flux contribution of the
NE component and $f^{\prime}$ is the area
covering factor of atomic gas for this component.
Assuming a spin temperature of 100\,K, which
is typical for atomic hydrogen gas in the Milky Way,
the column density of atomic gas is
$2.5 \times 10^{20} f{^{\prime}}^{-1}$\,cm$^{-2}$.
For the abundance ratios given in Table\,1
we set $f^{\prime} = 1$.

Using a typical value for $N({\rm H_2})/N({\rm CO})$
of $10^4$, the abundance ratio of molecular and atomic
hydrogen is $<0.25$ in the z$=$0.193 absorber.
This limit is not very illuminating, since the column
density of atomic gas is relatively low, corresponding
to an extinction $A_{\rm V} \approx 0.16$\,mag (e.g.
Savage et al. 1977). Since CO becomes self--shielded
only at $A_{\rm V} \ga 1$\,mag (e.g. Lucas \& Liszt 1994)
we do not expect a significant amount CO at these low
HI column densities. Molecular hydrogen, however, can
still be present, since it becomes self--shielded at
lower $N_{\rm HI}$.

\subsubsection{Limits to M$_{\rm H_2}$ from CO emission}

Our CO(1--0) spectra of the z$=$0.193 HI absorption can
also be used to derive an upper limit to the global molecular
mass content of any intervening galaxy at this distance.
The single dish telescope beam has a FWHP size of 52'' at
a frequency of 96.65\,GHz, corresponding to $\sim$143\,kpc
at z$=$0.193. Since atomic gas disks are considerably
smaller than this, an intervening galaxy causing the 21\,cm
HI absorption at z$=$0.193 must be within our telescope
beam.
The effective beam of the interferometer data is
6''$\times$2.4'' (16.5$\times$6.6 kpc)
 with a noise rms of 10\,mJy/beam.
The field of view is the same as for the single dish
observation, since 15--m telescopes were used in
both cases.

The single dish CO spectrum shown in Fig.\,4 has a
channel--to--channel noise rms of 1.0\,mK when binned
to a resolution of 17\,km\,s$^{-1}$. No emission is
evident in the spectrum and we derive a 3$\sigma$ upper
limit to the integrated CO intensity by assuming a
linewidth of 200\,km\,s$^{-1}$ to be $<$0.17\,K\,km\,s$^{-1}$.
The CO(1--0) spectrum does have a high frequency baseline
curvature, which has been removed in the spectrum shown
in Fig.\,4. The uncertainty in the derived integrated
intensity is limited by this curvature, but only for
emission more extended than 1000\,km\,s$^{-1}$. Since
such wide emission profiles are rarely encountered, we
believe that the baseline curvature is of little concern
in this case.
The interferometer data does not suffer from any baseline
curvature and gives a 3$\sigma$ upper limit to the integrated
CO intensity of $<$0.09\,K\,km\,s$^{-1}$

The CO line luminosity, expressed in K\,km\,s$^{-1}$\,pc$^2$,
can be written as
\begin{eqnarray}
L_{\rm CO}^{\prime} & = & \Omega_{s*b}\,D_{\rm L}^{2} {I_{\rm CO} \over
\left(1 + z\right)^3}\ \ {\rm K\ km\,s}^{-1}\ {\rm pc}^2\ \ ,
\end{eqnarray}
where $D_{\rm L}$ is the luminosity distance, $I_{\rm CO}$ is the
velocity integrated CO intensity expressed in K\,km\,s$^{-1}$,
and $\Omega_{s*b}$ is the solid angle of the source brightness
distribution convolved with the telescope beam, $\Omega_b$.
For distant sources and single dish telescopes, this convolution
is dominated by the solid angle of the beam and we can approximate
it with
$\Omega_{s*b} \approx \Omega_b \approx 26.56 \times B^2$,
(e.g. Solomon et al. 1997), where $B$ is the full width at half
power of the telescope beam measured in arcseconds\footnote{Solomon
et al. (1997) do not take the gaussian shape of the telescope beam
into account and their molecular mass formula therefore gives a
13\% lower mass estimate.}.
The mass of molecular gas is obtained by assuming a conversion
factor between CO luminosity and the column density of H$_2$:
$M_{\rm H_2} = \alpha\,L_{\rm CO}^{\prime}$. We use a standard
Milky Way conversion factor
$\alpha = 4.6$\,M$_{\odot}$\ (K\,km\,s$^{-1}$\ pc$^2$)$^{-1}$
(e.g. Solomon et al. 1997).
With these values our upper limits to $I_{\rm CO}$ transform
into a 3$\sigma$ upper limit to the H$_2$ mass of
$5 \times 10^{9}$\,M$_{\odot}$ and $2.5 \times 10^{9}$\,M$_{\odot}$
for the single dish and interferometer data, respectively.

\subsection{Additional absorption lines}

In addition to the 12 absorption lines presented in Wiklind
\& Combes (1996a) we have detected additional absorption
lines from CO(4$\leftarrow$3), H$_2$CO($2_{11}\leftarrow 1_{01}$)
and H$^{13}$CN(2$\leftarrow$1). The spectra are presented in
Fig.\,5 together with a new spectra of H$^{13}$CO$^+$(2$\leftarrow$1)
obtained with the IRAM 30--m telescope.

For the first time we detect one of the CO lines in this system.
The redshift of the absorber is such that the three first
rotational transitions of CO fall at frequencies unfavorable for
ground based observations. The CO(4$\leftarrow$3) line, however,
is shifted into the 1--mm band. We observed this line with
the IRAM 30--m telescope in May 1996 and the spectrum is presented
in Fig.\,5a. The line is rather strong, with a depth almost equal
to that of the saturated transitions of HCO$^+$ and HCN.
Using 40\% of the total flux and an excitation temperature of
6\,K, the CO column density is $2 \times 10^{18}$\,cm$^{-2}$.
Should $T_{\rm x}$ be higher, say 20\,K, the column density
becomes $0.2 \times 10^{18}$\,cm$^{-2}$. For higher excitation
temperatures the column density increases again.
The main uncertainty in this derivation is the continuum flux.
Due to unstable weather conditions it was not possible to
derive $T_{\rm c}$. We have used the continuum flux determined
from the 3-- and 2--mm bands, which corresponds to 1.2\,Jy and
converted this to antenna temperature in the 1--mm band.
Although indirect, the energy distribution of the radio
source PKS1830--211 is flat enough to make this approach
viable.
The excitation temperature of CO can be higher than the low
values found for HCO$^+$, HCN, N$_2$H$^+$, CS, etc. (see
Wiklind \& Combes 1996a). The reason is the lower electric
dipole moment of CO, making it more easily collisionally
excited and giving CO a higher level population at rotational
levels J$\geq$3 than do HCO$^+$ and HCN.
Different excitation temperatures for CO and the other molecular
species is consistent with the `weak LTE' conditions assumed
when deriving the excitation temperature, where it is assumed
that the level population of each molecule is characterized
by a unique temperature (that is: $T_{\rm x}=T_{\rm rot}$),
but this temperature can be different from one molecular
species to another (see Wiklind \& Combes 1995, 1997).

We also report the detection of ortho--H$_2$CO through 
the J$_{\rm K_{a}K_{b}} = 2_{11} \leftarrow 1_{01}$ line
(Fig.\,5b). The line appears to be non--saturated with
a maximum optical depth $\sim$1. The column density
of ortho--H$_2$CO is $2 \times 10^{15}$\,cm$^{-2}$,
assuming $T_{\rm x} = 6$\,K and a flux fraction of 40\%.

Absorption of H$^{13}$CO$^+$ was reported already in Wiklind \&
Combes (1996a), but the present spectra represents a
significant improvement to the signal--to--noise. The detection
of this isotopic species was one of the main arguments in favor
of a high H$^{12}$CO$^+$(2$\leftarrow$1) line opacity.
The column density of H$^{13}$CO$^+$, derived assuming the same
excitation temperature as for the main isotope of 6\,K, is
$2 \times 10^{13}$\,cm$^{-2}$. The H$^{12}$CO$^+$/H$^{13}$CO$^+$
ratio is thus $> 25$, possibly much higher.
From the similar absorption depth of HCO$^+$(2$\leftarrow$1) and
HCN(2$\leftarrow$1) we inferred that also the HCN line was heavily
saturated (Wiklind \& Combes 1996a). This is confirmed by the 
detection of the H$^{13}$CN(2$\leftarrow$1) line with the IRAM 30--m
telescope (Fig.\,5d). The column density of H$^{13}$CN is
$6 \times 10^{12}$\,cm$^{-2}$ (again using 40\% of the total flux
and adopting an excitation temperature of 6\,K).
The H$^{12}$CN/H$^{13}$CN ratio is $> 50$.

\section{Discussion}

\subsection{The lensing galaxy at z$=$0.886}

The detection of a second absorption line component at a redshift
z$=$0.886, shifted in velocity relative to the main one and
associated with the NE image of the background source,
opens up a unique possibility to derive the dynamical
mass of the central few kpc of the lensing galaxy.
This is valuable information when modelling the lens,
specifically when using time delay measurements to derive
cosmographic parameters such as the Hubble constant.
At the present this is hampered by the non--detection of
the lensing galaxy at optical and infrared wavelengths
(e.g. Djorgovski et al. 1992). 
This is likely to change with new and more powerful
infrared detectors.

It is, however, possible to use very simple but powerful
arguments, based on the assumption that the lensing
galaxy is a `normal' disk galaxy, to get reasonable ranges for
the mass of the lensing galaxy as well as the redshift
of the background radio source already from the present data.

The observed line of sight velocity in a disk galaxy of inclination
$i$, having a rotational velocity $V(R)$ is
$$
V_{\rm obs} = V(R) \cos{\theta} \sin{i}\ ,
$$
where $R$ is the galactocentric radius in the plane of
the galaxy and $\theta$ is the angle from the line of
nodes, as measured in the plane of the galaxy. $R$ and
$\theta$ are related to the radius $r$ and angle $\phi$ measured in
the plane of the sky through
$$
R = r \left(\cos{\phi}^{2} +
{\sin{\phi}^{2} \over \cos{i}^{2}}\right)^{1/2} ,\ \
\tan{\phi} = \tan{\theta} \cos{i}\ .
$$
If we assume (it will be justified below) that $i$ is small (i.e.
$i \la 30^{\circ}$), the differences between radii and
angles measured in the plane of the sky and those in
the plane of the galaxy are small ($\cos{i} \ga$ 0.87).
This is a reasonable assumption since, as we will show
below, the dynamical mass required to cause the image
separation of PKS1830--211 implies that the rotational
velocity of the lensing galaxy is more than twice larger than the
observed velocity separation of the main and satellite
absorption lines (e.g. $\sin{i} <0.5$, and $i < 30^{\circ}$).  
If we also make the assumption that the rotation curve is
similar in shape to those found in nearby galaxies,
we can assert that the rotational velocity is approximately
constant at radii $\ga 2$\,kpc.
This is seen from inspection of the rotation curves derived
for all different types of spiral galaxies by Rubin et al.
(1980, 1982, 1985).
Two different lens models have been produced for PKS1830--211
(Kochanek \& Narayan 1992; Nair et al. 1993). They both agree
that the center of the lensing galaxy is close to the center
of the ring seen in radio interferometer data and that the
lens has a small eccentricity. The model of
Nair et al. places the center of the lens somewhat closer
to the SW image than that of Kochanek \& Narayan. In the
former case the SW and NE cores are seen at projected
distances of 1.8\,kpc and 3.8\,kpc from the center.
In the model by Kochanek \& Narayan, both cores are 
$\sim$3\,kpc from the center of the lens.
In either case, it is reasonable to assume that the
rotational velocity of the lensing galaxy has reached a
relatively constant value at the projected location of
both the SW and NE cores; $V(R) \approx V_0$.
The observed velocity difference between two points in the
galaxy is then
$$
\Delta V_{\rm obs} = 
V_0 \left(\cos{\theta_1} - \cos{\theta_2}\right)\sin{i}\ ,
$$
where $\theta_j$ is the angle between the line of nodes and
the line joining the center of the lens with the point where
the line of sight crosses the disk, measured east of north.

The angles $\theta_j$ are 52$^{\circ}$ and 204$^{\circ}$
for the NE and SW components, respectively.
This gives a simple relation between the maximum velocity
$V_0$, the inclination $i$ and the observed
velocity separation of the absorption lines:
$|\Delta V| = V_0(1.5 \cos{\psi} + 1.2 \sin{\psi})\sin{i}$,
where $\psi$ is the position angle of the major axis of
the lensing galaxy. Geometrical considerations of the
lensing morphology implies that the position angle is
either $\sim 0^{\circ}$ or $\sim 90^{\circ}$ (Kochanek
\& Narayan 1992; Nair et al. 1993).
The mass inside a radius $R$ is given by the virial
theorem
$$
M(<R) \approx {\Delta V^{2} R \over \beta^{2} G \sin^{2}{i}}
\approx {7 \times 10^{9} \over \sin^{2}{i}}\ {\rm M_{\odot}}\ ,
$$
where $\beta$ is the factor depending on the position
angle of the lens: $\beta=1.5$ for $\psi = 0^{\circ}$
and $\beta=1.2$ for $\psi=90^{\circ}$.
We will adopt a position angle $\psi = 0^{\circ}$ in the
following, keeping in mind that values can change by at
most 20\% should the position angle be $90^{\circ}$.
In the last equality we used $\Delta V=147$\,km\,s$^{-1}$
and $R=3$\,kpc.

The two radio--components in PKS1830--211 have almost
identical substructure, suggesting that the components
are two parity--reversed images of the background quasar
caused by a nearly perfect alignment of the lens and
the source (e.g. Nair et al. 1993).
The separation of the NE and SW images can therefore to
first order be described as the effect of a single point
mass deflector of mass $M$, with the image separation
given by the `Einstein angle' $\Delta \theta_{\rm E}$
$$
\Delta \theta_{\rm E} = 4\sqrt{{GM \over c^2 D}} \approx
5.7 \times 10^{-6} \sqrt{{M \over D}}\ \ {\rm arcseconds}\ .
$$
In the last equality $M$ is expressed in solar mass and
the distance parameter $D$ in Gpc;
$D = D_{\rm d} D_{\rm s}/D_{\rm ds}$, where $D_{\rm d}$ is the
distance between the observer and the lens, $D_{\rm s}$ the
distance between observer and the source, and $D_{\rm ds}$
the distance between the lens and the source, all measured
as angular size distances (e.g. Schneider, Ehlers \& Falco 1992).
The last equation yields:
$$
M \approx 3 \times 10^{10}\,D\ \ {\rm M_{\odot}}\ , 
$$
($D$ in Gpc), and with the virial theorem:
\begin{eqnarray}
V_0 \approx 220 \sqrt{D}\ \ {\rm km\,s\,^{-1}}\ ,
\end{eqnarray}
In Fig.\,6 we plot $D$ as a function of the source
redshift for a fixed lens redshift of z$=$0.88582. The distances
are derived using H$_0 = 75$\,km\,s$^{-1}$\,Mpc$^{-1}$ and
for q$_0 = 0.5$ (full--drawn line) and q$_0 = 0.05$ (dotted line).
$D$ remains larger than 2 for all source redshift less than
z$\approx$7. This, together with the last equation,
means simply that the observed image separation of
$\sim$1" already requires a massive galaxy for the lens:
for a redshift of the background source $\ga 3$,
the required rotational velocity is 350\,km\,s$^{-1}$, 
corresponding to a massive early type spiral.
Lower redshifts for the background source would
require higher rotational velocities, corresponding to
an even more massive galaxy, which is not realistic for
spirals (unless several galaxies are considered to
contribute to the lensing).
Rotational velocities typical for late type spirals can
be excluded, since the redshift of the background source
becomes forbiddingly large.
Now, using the relation $\Delta V = 1.5 V_0 \sin{i}$ derived above,
with $\Delta V$ = 147\,km\,s$^{-1}$ and $V_0$ = 350\,km\,s$^{-1}$,   
yields an inclination of $i$ = 16$^\circ$.

How sensitively do these conclusions depend on the assumptions?
(1) The approximation of the lensing mass as a singular
point mass is only correct to first order if the lens
and the background source are aligned. This is
approximately true for PKS1830--211.
(2) The derivation of inclination assumes that the absorbing
gas is on circular orbits. If it has non--circular motion with
an amplitude comparable to the rotational velocity,
this simple approach is not valid.
However, the lens model of Kochanek \& Narayan (1993)
independently finds a low inclination.
(3) The presence of an absorber at z$=$0.193
introduces extra shear and, possibly, convergence of the
ray bundles. However, given the relatively simple image
configuration it is not likely that the extra mass
concentration at low redshift adds to the image
multiplicity. It can, however, influence the
interpretation of time delay measurements and Garrett
et al. (1996)
find from VLBI observations that the NE core component
has a more extended and complex morphology than the SW
component, which could be caused by the extra mass
at z$=$0.193.
Hence, our results regarding the mass of the lens at
z$=$0.886 and the lower limit to the redshift of the 
source are quite robust and we can make the
following postulates:

\begin{enumerate}

\item the lensing galaxy is an early
type spiral galaxy seen at a low inclination
($i < 20^{\circ}$).

\item The mass within a radius of $\sim$3\,kpc is in
the range $\sim 6-9 \times 10^{10}$\,M$_{\odot}$, which
is similar, within a factor 2, to the Milky Way galaxy.

\item The redshift of the background source is large,
most likely z$_{\rm s} > 3$.

\end{enumerate}

\subsection{The covering factor of the SW component}

Frye et al. (1997) observed PKS1830--211 with the BIMA millimeter
array and found HCO$^+$(2$\leftarrow$1) absorption only over the
SW component, as we do. However, the absorption line over this
region did not reach zero level, indicating a filling factor
of 70-80\%. This would mean that the prospect of using single
dish instruments to monitor the continuum and depth of saturated
absorption lines with the aim of deriving the time delay between
the NE and SW components can lead to ambiguities.
However, interferometric imaging of an object at low elevations
is difficult in the millimeter band and it is unclear how the
accuracy of the interferometer data can be assessed.
The flux ratio between the NE and SW continuum components in
the BIMA data is $1.14 \pm 0.05$\footnote{Taking the quoted
fluxes and errors from Frye et al. (1997), we get a flux ratio
of $1.14 \pm 0.24$.}
(Frye et al. 1997),
compatible with our value at the same epoch, but their total 
flux is much higher that what we find, either with Plateau
de Bure, IRAM--30m or SEST. In any case, the covering factor
is quite high, almost 1 for our measurements (see Table 1),
and is also compatible within 1 for the data of Frye et al (1997)
when their noise level is taken into account
or if their continuum level is overestimated by a factor
$\sim$1.4.
We do not think that this uncertainty is the main one in
the time--delay determination, the major uncertainty being
the continuum level measurement. 

It is not unexpected that a molecular cloud covers
entirely one of the PKS1830--211 image.
The angular extent of the continuum emission region in the
millimeter regime is not well known. The SW image remains unresolved
in 15\,GHz VLBA observations (Garrett et al. 1996) at a 
resolution of $0.6 \times 0.2$\,milliarcseconds (mas). The
NE component, however, is resolved with a complex structure.
An angular extent of 0.2\,mas corresponds to $\sim$1.1\,pc
but the size of the continuum source at millimeter wavelengths
can be considerably smaller. Millimeter VLBI of the
BL Lac 3C446 found an upper limit to the angular extent of
30\,$\mu$arcseconds (Lerner et al. 1993).

In Fig.\,7 we show the spectrum of the HCO$^+$(2$\leftarrow$1)
absorption obtained with the Plateau de Bure interferometer
with a velocity resolution of 0.5\,km\,s$^{-1}$.
The profile is asymmetric with an extension towards
negative velocities. This part of the profile is either
optically thin or has a covering factor less than the narrow
component. 
Saturated absorption lines of a finite width should be
flat--bottomed when viewed with a high velocity resolution.
The bottom of the narrow profile shown in Fig.\,7 appears
to be deeper at negative velocities than at positive.
The effect is only at the 1-2\% level. However, a similar
asymmetry is seen in the lower resolution
HCO$^+$(2$\leftarrow$1) spectra obtained from both the
IRAM 30--m and SEST telescopes (Fig.\,2). This asymmetry
could represent small differences in the covering factor
of the saturated absorption, corresponding to linear
dimensions of $\sim$5\% of the extent of the continuum source.

\section{Summary}

We have reported further molecular lines in absorption towards
PKS1830--211, and in particular CO(4$\leftarrow$3), the first
CO line. Molecular abundances are compatible with Milky Way
values.

We have also carried out interferometric measurements at three
different epochs, and found that the magnification ratio between
the two images has decreased from 1.8 to 1.0 during one year.
These measurements confirm that the main V=0 absorbing component
covers the SW image. We report about the detection of a
second component at V=--147.5\,km\,s$^{-1}$, which covers only
the NE image.

The presence of two absorption lines, one corresponding to the SW
image of the background source and the other to the NE component,
implies that the background radio source is situated at a redshift
$\ga 3$, and is consistent with the lensing galaxy being an early
type spiral seen almost face--on ($i\la 16^{\circ}$).
A first determination of the time delay of 44$\pm$9 days
has been reported by
van Ommen et al. (1995). Given the model dependence of this
measurement, it will be quite useful to try other
independent determinations at various frequencies. We confirm
here that with the SW component being nearly completely absorbed
at 3mm, single dish measurements are rapid and effective
for this experiment. It will however be necessary to confirm
from time to time that the filling factor is still $\approx$ 1
through interferometric measurement (but this has not varied
in the past year).

The lens geometry could be made more complex by the presence of
another galaxy on the line of sight, detected by HI absorption
by Lovell et al (1996) at z=0.1927.
We detect neither molecular absorption nor emission at this redshift;
the galaxy might not be exactly aligned with the quasar,
which would minimise its lensing effect. Alternatively, since
the HI absorption is seen only towards the NE image, and
molecular absorption has now been detected towards this component
also, this z=0.1927 absorption could be re--interpreted as the
absorption of a heavier molecule at z=0.88489, the measured
molecular redshift towards this NE component.
Exactly coinciding with this frequency we find the (7-1,0,7,6)
transition of the PS molecule, in a low enough excitation state.
This  interpretation is not likely, however, due to the
expected low abundance of PS in the ISM.

\acknowledgements
We thank the staff at SEST and IRAM (both at the 30--m telescope
and at the Plateau de Bure interferometer) for efficient help
and support during these observations.
TW acknowledges support from NFR (the Swedish Natural Science Research
Council) for this research and the European Southern Observatory for
hospitality during a Visitor Grant Programme in Santiago.

\clearpage


\clearpage

\begin{deluxetable}{lccc}
\small
\tablecaption{Magnification ratios at different epochs}
\tablewidth{0pt}
\tablehead{
\colhead{Date}                           &
\colhead{Total Flux}                     &
\colhead{NE/SW Ratio}                    &
\colhead{$\beta_{max}\tablenotemark{a}$}
}
\startdata
Oct 95 &  1.5     &  1.8 $\pm$ 0.1 &  1.     \nl
Apr 96 &  1.45    &  1.2 $\pm$ 0.1 &  0.9    \nl
Sep 96 &  0.93    &  1.0 $\pm$ 0.1 &  1.     \nl
\nl

\enddata
\tablenotetext{a}{The maximum covering factor of the SW component.}
\end{deluxetable}

\clearpage

\begin{deluxetable}{llrrcrrrr}
\footnotesize
\tablecaption{Non--detection of absorption at z$=$0.19267}
\tablewidth{0pt}
\tablehead{
\colhead{Molecule}                               &
\colhead{Transition}                             &
\colhead{$\nu_{\rm rest}$\tablenotemark{a}}      &
\colhead{$\nu_{\rm obs}$\tablenotemark{b}}       &
\colhead{T$_{\rm cont}$\tablenotemark{c}}        &
\colhead{ }                                      &
\colhead{N$_{\rm tot}$\ [cm$^{-2}$]}             &
\colhead{ }                                      &
\colhead{N/N$_{\rm HI}$\tablenotemark{d}}        \nl
\nl
\colhead{ }                                      &
\colhead{ }                                      &
\colhead{GHz}                                    &
\colhead{GHz}                                    &
\colhead{mK}                                     &
\colhead{100\%}                                  &
\colhead{60\%}                                   &
\colhead{40\%}                                   &
\colhead{ }                                      
}
\startdata
CO       & 1$\leftarrow$0 & 115.271204 &  96.650 & 50 &
$<3 \times 10^{15}$ & $<6 \times 10^{15}$ & $< 1 \times 10^{16}$ & $< 2 \times 10^{-5}$ \nl
\nl
HCO$^+$  & 2$\leftarrow$1 & 178.375065 & 149.559 & 26 &
$<3 \times 10^{12}$ & $<5 \times 10^{12}$ & $< 8 \times 10^{12}$ & $< 2 \times 10^{-8}$ \nl
\nl

\enddata
\tablenotetext{a}{The rest--frequency of the observed molecules taken from Lovas (1992).}
\tablenotetext{b}{Derived from $\nu_{\rm obs} = \nu_{\rm rest}/(1+z_{\rm a})$.}
\tablenotetext{c}{Continuum level measured in $T_{\rm A}^{*}$.}
\tablenotetext{d}{Ratio of molecular column density and HI (see text for details).}
\end{deluxetable}

\clearpage

\begin{figure}
\epsscale{1.0}
\plotone{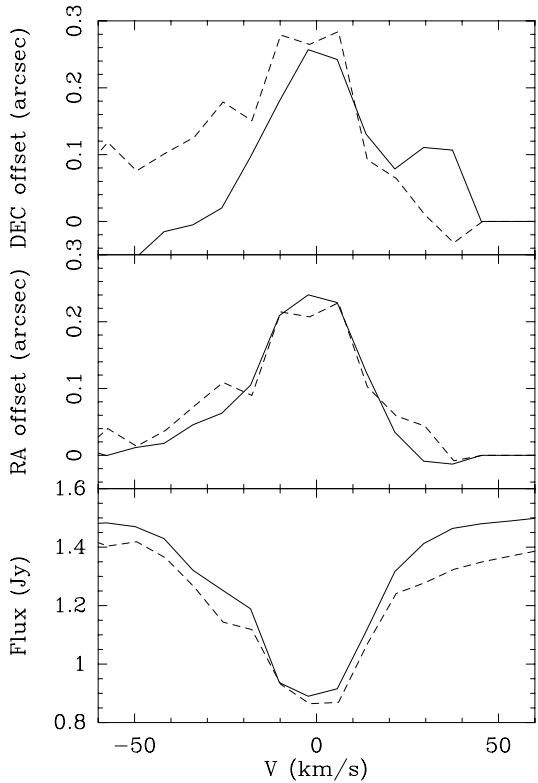}
\caption{HCO$^+$(2$\leftarrow$1) spectra
(bottom) obtained with the IRAM Plateau de Bure interferometer
in September--October 1995 (dashed line) and April 1996 (full line).
The velocity resolution is 7.9\,km\,s$^{-1}$ in both cases.
Also shown is  the relative location of the phase--center in RA
and DEC (top and middle).\label{pks1830-fig1}}
\end{figure}


\begin{figure}
\epsscale{0.8}
\plotone{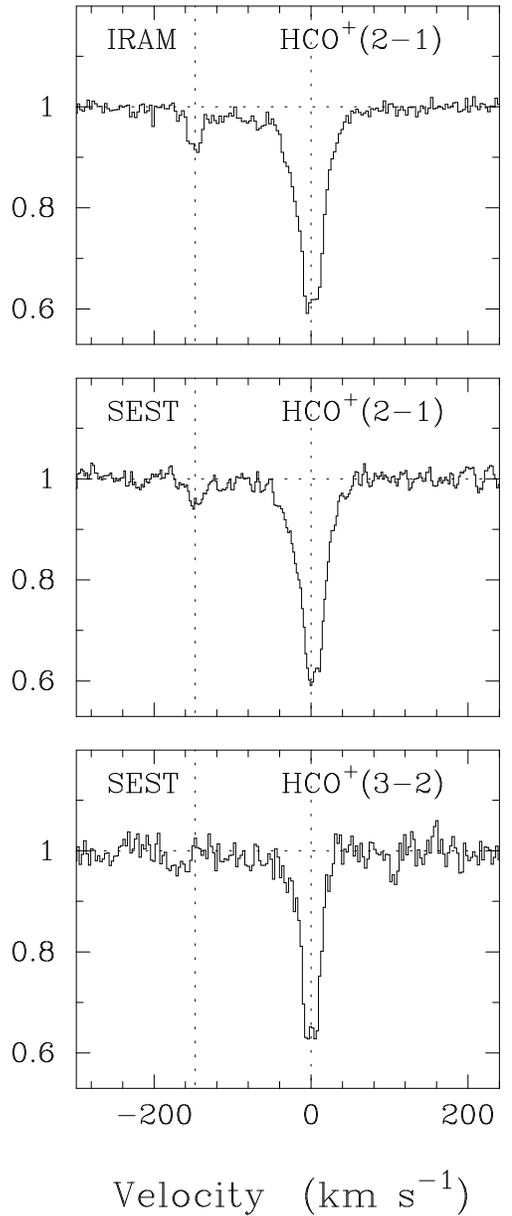}
\caption{The HCO$^{+}$(2$\leftarrow$1)
absorption profile from the 30--m IRAM (top) and the 15--m
SEST (middle) telescopes, showing the second absorption
component at $-147.5$\,km\,s$^{-1}$. The continuum level
has been normalized to unity. The vertical dotted lines
mark velocities at $-147.5$\,km\,s$^{-1}$ and 0\,km\,s$^{-1}$.
In the bottom, we show the HCO$^+$(3$\leftarrow$2) spectrum,
obtained with the SEST.
No line at $-147.5$\,km\,s$^{-1}$ can be seen in the J$=$3--2
line. The velocity resolution is 3.2, 2.2 and 2.9\,km\,s$^{-1}$
from top to bottom.\label{pks1830-fig2}}
\end{figure}

\clearpage

\begin{figure}
\epsscale{1.0}
\plotone{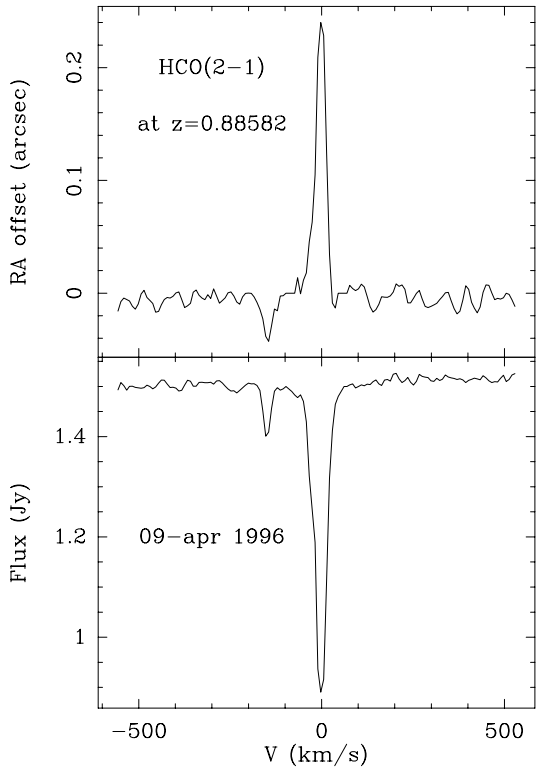}
\caption{Results of the fit of visibility data from
April 1996 observations: the integrated spectrum is at bottom, and the
phase--center position in right ascension at top: this shows that the
$V=-147.5$\,km\,s$^{-1}$ component is absorbed in front of the NE image.
\label{pks1830-fig3}}
\end{figure}


\begin{figure}
\epsscale{0.7}
\plotone{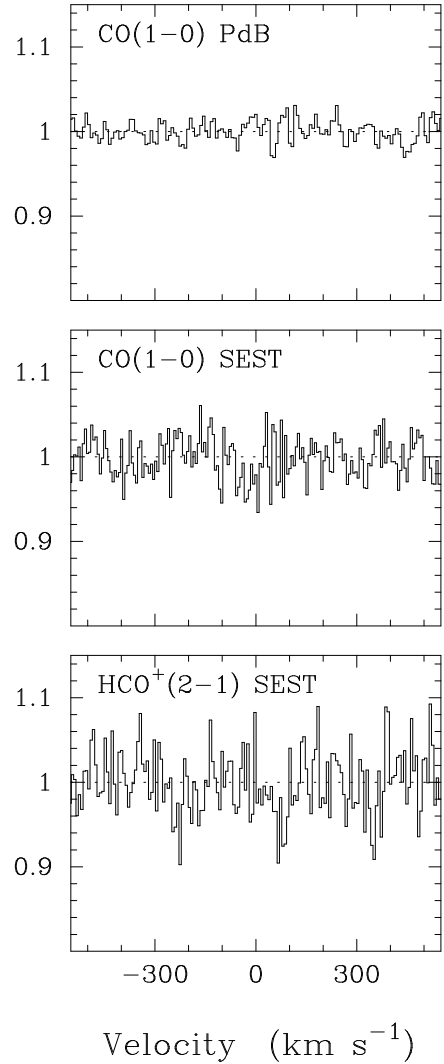}
\caption{The CO(1--0) and HCO$^+$(2--1) spectra
from the z$=$0.19267 HI absorption system. No absorption, nor emission
is present. The velocity resolution is 7.8 and 10.5\,km\,s$^{-1}$ for
the CO spectra  and 11.1\,km\,s$^{-1}$ for HCO$^+$. The velocity scale
is heliocentric and the SEST spectra have been normalized to the continuum
levels of 50\,mK and 26\,mK for CO and HCO$^+$, respectively.\label{pks1830-fig4}}
\end{figure}

\clearpage

\begin{figure}
\epsscale{1.0}
\plotone{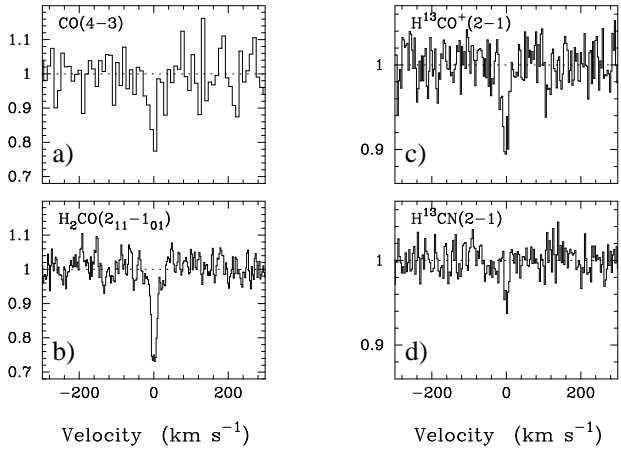}
\caption{
{\bf a)}CO(4$\leftarrow$3),
{\bf b)} H$_2$CO(2$_{11}\leftarrow$1$_{01}$),
{\bf c)} H$^{13}$CO$^+$(2$\leftarrow$1) and
{\bf d)}H$^{13}$CN(2$\leftarrow$1)
absorption at z$=$0.88582. All spectra except H$_2$CO,
have been obtained with the IRAM 30--m telescope. H$_2$CO
was observed with the 15--m SEST. The velocity resolution
is 9.2\,km\,s$^{-1}$ for the CO spectrum,
2.6\,km\,s$^{-1}$ for the H$_2$CO spectrum and
3.25\,km\,s$^{-1}$ for the H$^{13}$CO$^+$ and H$^{13}$CN spectra.
The total continuum level has been normalized to unity and
the velocity scale is heliocentric in the restframe of the
z$=$0.88582 absorber.\label{pks1830-fig5}}
\end{figure}


\begin{figure}
\epsscale{1.0}
\plotone{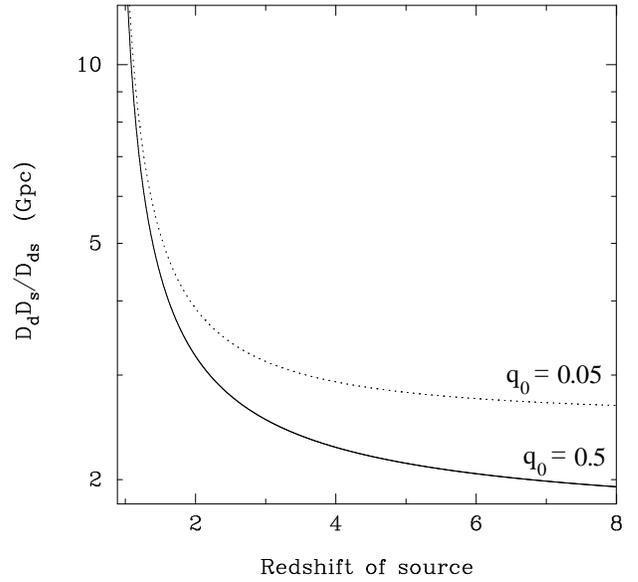}
\caption{The distance parameter $D = D_{\rm d} D_{\rm s}/D_{\rm ds}$,
where $D_{\rm d}$ is the angular size distance between the observer and
the deflector, $D_{\rm s}$ the distance between the observer and the source,
and $D_{\rm ds}$ is the distance between the deflector and the source. In
this plot the redshift of the deflector is held fixed at z$=$0.88582. 
The relation is given for a Hubble parameter
H$_0 = 75$\,km\,s$^{-1}$\,Mpc$^{-1}$ and for q$_0 = 0.5$ and
q$_0 = 0.05$.\label{pks1830-fig6}}
\end{figure}

\clearpage

\begin{figure}
\epsscale{1.3}
\plotone{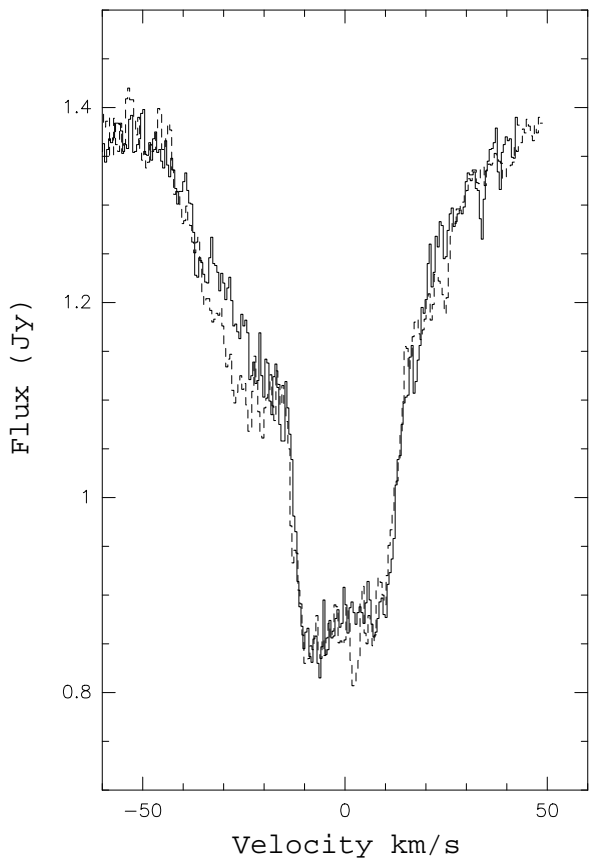}
\caption{High spectral resolution (0.5\,km\,s$^{-1}$)
Plateau de Bure  absorption profile in HCO+(2$\leftarrow$1) in April 96
(full line) and September--October 1995 (dashed line).\label{pks1830-fig7}}
\end{figure}

\end{document}